\documentclass[12pt]{article}
\usepackage{amsmath,amsfonts,amssymb}

\makeatletter
\def\numberbysection{\@addtoreset{equation}{section}
    \def\theequation{\thesection.\arabic{equation}}}
\makeatother

\numberbysection

\newcommand{\be}{\begin{eqnarray}}
\newcommand{\ee}{\end{eqnarray}}
\newcommand{\non}{\nonumber}

\newcommand{\tr}{\mathop{\rm tr}\nolimits}
\newcommand{\str}{\mathop{\rm str}\nolimits}

\begin{document}

\begin{titlepage}
\strut\hfill UMTG--260
\vspace{.5in}
\begin{center}

\LARGE An alternative $S$-matrix \\
for ${\cal N}=6$ Chern-Simons theory ?\\
\vspace{1in}
\large Changrim Ahn \footnote{
       Department of Physics, Ewha Womans University,
       Seoul 120-750, South Korea; ahn@ewha.ac.kr} and
       Rafael I. Nepomechie \footnote{
       Physics Department, P.O. Box 248046, University of Miami,
       Coral Gables, FL 33124 USA;  nepomechie@physics.miami.edu}\\

\end{center}

\vspace{.5in}

\begin{abstract}
We have recently proposed an $S$-matrix for the planar limit of the
${\cal N}=6$ superconformal Chern-Simons theory of Aharony, Bergman,
Jafferis and Maldacena which leads to the all-loop Bethe ansatz
equations conjectured by Gromov and Vieira.  An unusual feature of
this proposal is that the scattering of $A$ and $B$ particles is
reflectionless.  We consider here an alternative $S$-matrix, for which
$A-B$ scattering is not reflectionless.  We argue that this $S$-matrix
does not lead to the Bethe ansatz equations which are consistent with
perturbative computations.
\end{abstract}

\end{titlepage}

\setcounter{footnote}{0}

\section{Introduction}\label{sec:intro}

The fact that the 3-dimensional ${\cal N}=6$ superconformal
Chern-Simons (CS) theory of Aharony, Bergman, Jafferis and Maldacena
\cite{ABJM} has a planar limit suggests that it may have further
features in common with 4-dimensional ${\cal N}=4$ superconformal
Yang-Mills (YM) theory.  Indeed, it was shown by Minahan and Zarembo
\cite{MZ} (see also \cite{BR}) that the two-loop anomalous dimensions
of the scalar operators in planar ${\cal N}=6$ CS theory are described
by a certain integrable spin chain.  Furthermore, they conjectured
two-loop Bethe ansatz equations (BAEs) for the full theory.  Gromov
and Vieira \cite{GV} subsequently conjectured all-loop BAEs, which
reduce to those of Minahan and Zarembo in the weak-coupling limit.
Recently, three groups \cite{MR, AAB, Kr} computed the one-loop
correction to the energy of a folded spinning string, and seemed to
find disagreement with the prediction of the all-loop BAEs.  This
controversy may be resolved by a non-zero one-loop correction in the
central interpolating function $h(\lambda)$ as suggested recently in
\cite{MRT}.  (See also \cite{GM}.)

Based on the spectrum and symmetries of the model \cite{MZ, NT, GGY,
GHO}, we proposed an all-loop $S$-matrix \cite{AN} which reproduces the
all-loop BAEs.  That $S$-matrix has the unusual feature that the
scattering of $A$ and $B$ particles is reflectionless,
\be 
A\, B \rightarrow B\, A \non 
\ee 
(instead of $A\, B \rightarrow B\, A + A\, B$).  Given the uncertainty
in these all-loop proposals, one may well wonder whether there exists
another $S$-matrix which 
\begin{enumerate}
    \item[(i)] is not reflectionless; and 
    \item[(ii)] is consistent with the two-loop BAEs of Minahan and Zarembo \cite{MZ},
which are on firmer ground.
\end{enumerate}
This note is an effort to address this question.  Unfortunately, we do
not give a definitive answer.  Nevertheless, our failure to find such
an alternative $S$-matrix provides increased confidence in our
original proposal \cite{AN}, and in the corresponding all-loop BAEs
\cite{GV}.

The outline of this paper is as follows.  In Section \ref{sec:Smatrix}
we construct a candidate alternative $S$-matrix.  The key feature of
this $S$-matrix which allows for reflection is that it factorizes into
the product of a nontrivial flavor part and an $SU(2|2)$ part.  In
order to simplify the ensuing analysis, we make the unphysical
assumption that the flavor part is $SU(2)$-invariant.  (We later argue
that this simplifying assumption does not alter the main conclusion.)
In Section \ref{sec:ABA} we derive the corresponding all-loops BAEs by
diagonalizing the Bethe-Yang matrix.  We perform the weak-coupling
limit, and show that the result is not consistent with the two-loop
BAEs \cite{MZ}.  We conclude in Section \ref{sec:discussion} with a
brief discussion of our results.

\section{$S$-matrix}\label{sec:Smatrix}

We represent the elementary excitations by Zamolodchikov-Faddeev
operators $A_{a\, i}^{\dagger}(p)$, where $a \in \{1, 2\}$ is a
flavor index ($a=1$ corresponds to an $A$-particle, and $a=2$
corresponds to a $B$-particle), and $i \in \{1, 2, 3, 4\}$ is the
$SU(2|2)$ index.  When acting on the vacuum state $|0\rangle$, these
operators create corresponding asymptotic particle states
of momentum $p$ and energy $E$ given by \cite{NT, GGY, GHO, BT}
\be
E = \sqrt{\frac{1}{4}+4 g^{2} \sin^{2}\frac{p}{2}} \,,
\ee
where $g$ is a function of the 't Hooft coupling
\be
g=h(\lambda) \,,
\label{gvalue}
\ee
with $h(\lambda) \sim \lambda$ for small $\lambda$, and
$h(\lambda) \sim \sqrt{\lambda/2}$ for large $\lambda$.

A way to allow for reflection of $A$ and $B$ particles, while still
maintaining integrability, is to assume that the $S$-matrix factorizes
into the product of a nontrivial flavor part and an $SU(2|2)$ part,
\be
A_{a\, i}^{\dagger}(p_{1})\, A_{b\, j}^{\dagger}(p_{2}) =
S_{0}(p_{1}, p_{2})\, S_{a\, b}^{a'\, b'}(p_{1}, p_{2})\,
\widehat S_{i\, j}^{i'\, j'}(p_{1}, p_{2})\, A_{b'\, j'}^{\dagger}(p_{2})\,
A_{a'\, i'}^{\dagger}(p_{1})\,,
\label{Smatrix}
\ee
where both the flavor $S$-matrix $S_{a\, b}^{a'\,
b'}(p_{1}, p_{2})$ and the $SU(2|2)$ $S$-matrix $\widehat S_{i\, 
j}^{i'\, j'}(p_{1}, p_{2})$ satisfy the Yang-Baxter equation (YBE),
and $S_{0}(p_{1}, p_{2})$ is an unknown scalar factor.

The $SU(2|2)$ part is essentially fixed \cite{St,Be1}, with with $g$
given by (\ref{gvalue}).  More precisely, in order to carry out the
asymptotic Bethe ansatz analysis below, we assume that $\widehat
S_{i\, j}^{i'\, j'}(p_{1}, p_{2})$ is the graded version \cite{MM} of
the $SU(2|2)$-invariant $S$-matrix given in \cite{AFZ}.

Since the only known symmetry relating $A$ and $B$ particles is $CP$
symmetry, the flavor $S$-matrix should not have any more symmetry.
Solutions of the YBE with only discrete symmetry are known, such as
the $R$-matrix of the XYZ spin chain/8-vertex model; and in principle,
we could proceed by assuming that the flavor $S$-matrix is of that
form.  However, in order to simplify the ensuing analysis, we instead
make the unphysical assumption that the flavor $S$-matrix is
$SU(2)$-invariant.  We shall later argue that this simplifying
assumption does not alter the main conclusion.

As is well-known (see, e.g., \cite{Be2}), $SU(2)$ symmetry and
factorizability almost completely fix the structure of the $S$-matrix.
Indeed, $SU(2)$ symmetry implies that, up to an overall scalar factor,
\be
S_{a\, b}^{a'\, b'}(p_{1}, p_{2}) = i \delta_{a}^{b'} \delta_{b}^{a'}
+ f(p_{1}, p_{2}) \delta_{a}^{a'} \delta_{b}^{b'} \,,
\ee
where $f(p_{1}, p_{2})$ is an arbitrary scalar function of $p_{1},
p_{2}$. The YBE
\be
S_{12}(p_{1}, p_{2})\, S_{13}(p_{1}, p_{3})\, S_{23}(p_{2}, p_{3})\ =
S_{23}(p_{2}, p_{3})\, S_{13}(p_{1}, p_{3})\, S_{12}(p_{1}, p_{2})
\label{YBE}
\ee
then implies that
\be
f(p_{1}, p_{2}) = f(p_{1}, p_{3}) - f(p_{2}, p_{3}) \,,
\ee
which in turn implies that
\be
f(p_{1}, p_{2}) = \alpha(p_{1}) -  \alpha(p_{2}) \,,
\ee
where $\alpha(p)$ is an arbitrary function of $p$. In the
weak-coupling limit, $\alpha(p)$ must be a linear function of $p$, say
\be
\alpha(p) = p \,, \quad \mbox{(weak coupling)}
\label{weak}
\ee
in order that $f(p_{1}, p_{2})$ be a function of $p_{1} - p_{2}$,
i.e., that the $S$-matrix have the ``difference'' property. We
conclude that
\be
S_{a\, b}^{a'\, b'}(p_{1}, p_{2}) = i \delta_{a}^{b'} \delta_{b}^{a'}
+ \left(  \alpha(p_{1}) -  \alpha(p_{2})  \right)
\delta_{a}^{a'} \delta_{b}^{b'} \,.
\label{su2S}
\ee
In matrix form,
\be
S(p_{1}, p_{2}) =  \left( \begin{array}{cccc}
a(p_{1}, p_{2}) & 0 & 0 & 0 \\
0 & b(p_{1}, p_{2}) & i & 0 \\
0 & i & b(p_{1}, p_{2}) & 0 \\
0 & 0 & 0 & a(p_{1}, p_{2})
\end{array} \right) \,,
\label{su2Smatrix}
\ee
where
\be
a(p_{1}, p_{2}) = \alpha(p_{1}) -  \alpha(p_{2}) + i \,, \qquad
b(p_{1}, p_{2}) = \alpha(p_{1}) -  \alpha(p_{2}) \,.
\ee
Note that
\be
b(p_{1}, p_{2}) = -b(p_{2}, p_{1}) \,.
\label{bantisym}
\ee

The $S$-matrix (\ref{Smatrix}), unlike the one which we proposed in
\cite{AN}, does allow for reflection in $A-B$ scattering.  Examples of
integrable models with $S$-matrices of product form include
\cite{product}.  To determine the function $\alpha$, one may need to
impose CP symmetry between $A$- and $B$-particles
which leads to a crossing relation.  We shall not pursue this here
since our conclusion does not depend on the explicit form of $\alpha$.

\section{Asymptotic Bethe ansatz}\label{sec:ABA}

We now proceed to derive the corresponding all-loop BAEs.  The analysis
is similar to the one for ${\cal N}=4$ YM theory \cite{Be1, MM};
and as in \cite{AN}, we follow closely the latter reference.  We
consider a set of $N$ particles with momenta $p_{i}$ ($i= 1, \ldots,
N$) which are widely separated on a ring of length $L'$.  Quantization
conditions for these momenta follow from imposing periodic boundary
conditions on the wavefunction.  Taking a particle with momentum
$p_{k}$ around the ring leads to the Bethe-Yang equations
\be
e^{-i p_{k} L'} = \Lambda( \lambda=p_{k}, \{ p_{i} \}; \{\lambda_{j},
\mu_{j}, \xi_{j} \})\,, \quad k = 1, \ldots, N \,,
\label{BetheYang}
\ee
where $\Lambda( \lambda, \{ p_{i} \}; \{\lambda_{j},
\mu_{j}, \xi_{j} \})$ are the
eigenvalues of the transfer matrix
\be
t(\lambda, \{ p_{i} \}) =
\Lambda_{0}(\lambda, \{ p_{i} \})\,
t_{SU(2)}(\lambda, \{ p_{i} \}) \otimes
t_{SU(2|2)}(\lambda, \{ p_{i} \}) \,,
\ee
where
\be
\Lambda_{0}(\lambda, \{ p_{i} \})  &=& \prod_{i=1}^{N}S_{0}(\lambda, p_{i})\,, \non \\
t_{SU(2)}(\lambda, \{ p_{i} \})  &=&
\tr_{a} S_{a\, 1}(\lambda, p_{1}) \cdots S_{a\, N}(\lambda, p_{N})
\,, \non \\
t_{SU(2|2)}(\lambda, \{ p_{i} \})  &=&
\str_{a}
\widehat S_{a\, 1}(\lambda, p_{1}) \cdots
\widehat S_{a\, N}(\lambda, p_{N})\,.
\ee
Hence, the eigenvalues are given by
\be
\Lambda( \lambda, \{ p_{i} \}; \{\lambda_{j},
\mu_{j}, \xi_{j} \}) =
\Lambda_{0}(\lambda, \{ p_{i} \})\,
\Lambda_{SU(2)}(\lambda, \{ p_{i} \}; \{\xi_{j} \})\,
\Lambda_{SU(2|2)}(\lambda, \{ p_{i} \}; \{\lambda_{j}, \mu_{j}\}) \,,
\ee
where the $SU(2)$ part is given by the well-known algebraic Bethe
ansatz result
\be
\Lambda_{SU(2)}(\lambda, \{ p_{i} \}; \{\xi_{j} \}) =
\prod_{i=1}^{N}a(\lambda, p_{i}) \prod_{j=1}^{m_{0}}s(\xi_{j}\,,
\lambda) +
\prod_{i=1}^{N}b(\lambda, p_{i}) \prod_{j=1}^{m_{0}}s(\lambda\,,
\xi_{j}) \,,
\ee
with
\be
s(p_{1}, p_{2}) = \frac{a(p_{1}, p_{2})}{b(p_{1}, p_{2})} =
\frac{\alpha(p_{1})-\alpha(p_{2})+i}{\alpha(p_{1})-\alpha(p_{2})} \,,
\ee
and $\{ \xi_{j} \}$ obey the BAEs
\be
\prod_{i=1}^{N} s(\xi_{k}, p_{i}) = \prod_{j=1 \atop j\ne k}^{m_{0}}
\frac{s(\xi_{k}, \xi_{j})}{s(\xi_{j}, \xi_{k})} \,, \qquad k = 1\,,
\ldots \,, m_{0} \,.
\label{BAEsu2}
\ee
In particular, due to the property (\ref{bantisym}), the eigenvalues
at $\lambda=p_{k}$ are given by
\be
\Lambda_{SU(2)}(\lambda=p_{k}, \{ p_{i} \}; \{\xi_{j} \}) =
\prod_{i=1}^{N}a(p_{k}, p_{i}) \prod_{j=1}^{m_{0}}s(\xi_{j}\,, p_{k})
\,. \label{su2part}
\ee
Moreover, the $SU(2|2)$ part is given by \cite{MM}
\be
\lefteqn{\Lambda_{SU(2|2)}(\lambda, \{ p_{i} \}; \{\lambda_{j}, \mu_{j}\})
 =\prod_{i=1}^{N}\left[ \frac{x^{+}(\lambda) - x^{-}(p_{i})}
{x^{-}(\lambda) - x^{+}(p_{i})}
\frac{\eta(p_{i})}{\eta(\lambda)} \right]
\prod_{j=1}^{m_{1}}\left[\eta(\lambda)
\frac{x^{-}(\lambda)-x^{+}(\lambda_{j})}
{x^{+}(\lambda)-x^{+}(\lambda_{j})}\right]}\non \\
&&-
\prod_{i=1}^{N}\left[ \frac{x^{+}(\lambda) - x^{+}(p_{i})}
{x^{-}(\lambda) - x^{+}(p_{i})}
\frac{1}{\eta(\lambda)} \right]
\Bigg\{
\prod_{j=1}^{m_{1}}\left[ \eta(\lambda)
\frac{x^{-}(\lambda)-x^{+}(\lambda_{j})}
{x^{+}(\lambda)-x^{+}(\lambda_{j})}\right]
\prod_{l=1}^{m_{2}}
\frac{x^{+}(\lambda) + \frac{1}{x^{+}(\lambda)} - \tilde \mu_{l} +
\frac{i}{2g}}
{x^{+}(\lambda) + \frac{1}{x^{+}(\lambda)} - \tilde \mu_{l} - \frac{i}{2g}}\non \\
&&+\prod_{j=1}^{m_{1}}\left[ \eta(\lambda)
\frac{x^{+}(\lambda_{j})-\frac{1}{x^{+}(\lambda)}}
{x^{+}(\lambda_{j})-\frac{1}{x^{-}(\lambda)}}\right]
\prod_{l=1}^{m_{2}}
\frac{x^{-}(\lambda) + \frac{1}{x^{-}(\lambda)} - \tilde \mu_{l} -
\frac{i}{2g}}
{x^{-}(\lambda) + \frac{1}{x^{-}(\lambda)} - \tilde \mu_{l} + \frac{i}{2g}} \Bigg\} \non \\
&&+
\prod_{i=1}^{N}\left[ \frac{x^{+}(\lambda) - x^{+}(p_{i})}
{x^{-}(\lambda) - x^{+}(p_{i})}
\frac{1-\frac{1}{x^{-}(\lambda) x^{+}(p_{i})}}
{1-\frac{1}{x^{-}(\lambda) x^{-}(p_{i})}}
\frac{\eta(p_{i})}{\eta(\lambda)} \right]
\prod_{j=1}^{m_{1}}\left[ \eta(\lambda)
\frac{x^{+}(\lambda_{j})-\frac{1}{x^{+}(\lambda)}}
{x^{+}(\lambda_{j})-\frac{1}{x^{-}(\lambda)}}\right] \,,
\label{thateigenvalues}
\ee
where $\eta(\lambda) = e^{i \lambda/2}$, and
the corresponding BAEs are given by
\be
e^{i P/2}
\prod_{i=1}^{N} \frac{x^{+}(\lambda_{j}) - x^{-}(p_{i})}
{x^{+}(\lambda_{j}) - x^{+}(p_{i})}
&=& \prod_{l=1}^{m_{2}}
\frac{x^{+}(\lambda_{j}) + \frac{1}{x^{+}(\lambda_{j})}
- \tilde \mu_{l} + \frac{i}{2g}}
{x^{+}(\lambda_{j}) + \frac{1}{x^{+}(\lambda_{j})}
- \tilde \mu_{l} - \frac{i}{2g}} \,,  \qquad j = 1, \ldots, m_{1} \,, \non \\
\prod_{j=1}^{m_{1}}\frac{\tilde \mu_{l} -
x^{+}(\lambda_{j})-\frac{1}{x^{+}(\lambda_{j})}+\frac{i}{2g}}
{\tilde \mu_{l} -
x^{+}(\lambda_{j})-\frac{1}{x^{+}(\lambda_{j})}-\frac{i}{2g}}
&=&
\prod_{k=1 \atop k\ne l}^{m_{2}}
\frac{\tilde \mu_{l} - \tilde \mu_{k} + \frac{i}{g}}
{\tilde \mu_{l} - \tilde \mu_{k} - \frac{i}{g}} \,, \qquad l = 1, \ldots,
m_{2} \,,
\label{BAEsu22}
\ee
where
\be
\frac{x^{+}(\lambda)}{x^{-}(\lambda)} = e^{i \lambda}\,, \quad
x^{+}(\lambda)+\frac{1}{x^{+}(\lambda)}
-x^{-}(\lambda)-\frac{1}{x^{-}(\lambda)} = \frac{i}{g}\,, \qquad
P=\sum_{i=1}^{N}p_{i}  \,.
\ee
In particular, the eigenvalue at $\lambda=p_{k}$ is given simply by
\be
\Lambda_{SU(2|2)}(\lambda=p_{k}, \{ p_{i} \}; \{\lambda_{j}, \mu_{j}\})
= \prod_{i=1}^{N}\left[ \frac{x^{+}(p_{k}) - x^{-}(p_{i})}
{x^{-}(p_{k}) - x^{+}(p_{i})}
\frac{\eta(p_{i})}{\eta(p_{k})} \right]
\prod_{j=1}^{m_{1}}\left[\eta(p_{k})
\frac{x^{-}(p_{k})-x^{+}(\lambda_{j})}
{x^{+}(p_{k})-x^{+}(\lambda_{j})}\right]\,. \non\\
\label{su22part}
\ee

In view of (\ref{su2part}), (\ref{su22part}), the Bethe-Yang equations
(\ref{BetheYang}) take the form
\be
e^{i p_{k} \left(-L'+\frac{N}{2}-\frac{m_{1}}{2}\right)} &=& e^{i P/2}
\prod_{i=1}^{N}\left\{ S_{0}(p_{k}, p_{i})\, a(p_{k}, p_{i})
\left[ \frac{x^{+}(p_{k}) - x^{-}(p_{i})}
{x^{-}(p_{k}) - x^{+}(p_{i})} \right] \right\} \non \\
& & \times
\prod_{j=1}^{m_{0}}s(\xi_{j}\,, p_{k})
\prod_{j=1}^{m_{1}}
\frac{x^{-}(p_{k})-x^{+}(\lambda_{j})}
{x^{+}(p_{k})-x^{+}(\lambda_{j})}
\,, \quad k = 1, \ldots, N \,,
\label{BetheYang2}
\ee
where $\{\lambda_{j}, \mu_{j}, \xi_{j} \}$ are determined by the BAEs
(\ref{BAEsu2}), (\ref{BAEsu22}).

Following \cite{MM, AN}, we make the identifications
\be
x^{\pm}(p_{k}) &=&x^{\pm}_{4, k}\,, \quad k = 1, \ldots,
K_{4} \equiv N \,, \non \\
x^{+}(\lambda_{j}) &=& \frac{1}{x_{1, j}}\,, \quad j = 1, \ldots, K_{1}\,,
\non \\
x^{+}(\lambda_{K_{1}+j}) &=& x_{3, j}\,, \quad j = 1, \ldots,
K_{3}\,, \quad K_{1} + K_{3} \equiv m_{1} \,, \non \\
\tilde \mu_{j} &=& \frac{u_{2, j}}{g}\,, \quad j = 1, \ldots, K_{2}
\equiv m_{2} \,,
\ee
and also define
\be
u_{4, j} = x^{+}_{4, j} + \frac{1}{x^{+}_{4, j}} - \frac{i}{2} =
x^{-}_{4, j} + \frac{1}{x^{-}_{4, j}} + \frac{i}{2} \,,
\label{ufourj}
\ee
and
$u_{i, j} = g\left(x_{i, j} + \frac{1}{x_{i, j}}\right)$ for $i=1, 3$.
We assume the zero-momentum condition
\be
P = \sum_{j=1}^{K_{4}} p_{4, j} =0 \,,
\ee
and (for aesthetic reasons) we perform the shift
\be
\alpha(\xi_{j}) \rightarrow \alpha(\xi_{j}) - \frac{i}{2} \,.
\ee
The Bethe-Yang equations (\ref{BetheYang2}) become
\be
\lefteqn{e^{i p_{4,k} \left(-L'+\frac{K_{4}+K_{1}-K_{3}}{2}\right)} =
\prod_{i=1}^{K_{4}}\left\{ S_{0}(p_{4,k}, p_{4,i})\,
\left[\alpha(p_{4,k}) -\alpha(p_{4,i}) + i\right]
\left(\frac{x^{+}_{4,k} - x^{-}_{4,i}}
{x^{-}_{4,k} - x^{+}_{4,i}} \right) \right\} }\non \\
& & \times
\prod_{j=1}^{m_{0}}\frac{\alpha(\xi_{j}) -\alpha(p_{4,k}) + \frac{i}{2}}
{\alpha(\xi_{j}) -\alpha(p_{4,k}) - \frac{i}{2}}
\prod_{j=1}^{K_{1}}
\frac{1 - \frac{1}{x^{-}_{4,k} x_{1,j}}}
{1 - \frac{1}{x^{+}_{4,k} x_{1,j}}}
\prod_{j=1}^{K_{3}}
\frac{x^{-}_{4,k}-x_{3,j}}
{x^{+}_{4,k}-x_{3,j}}
\,, \quad k = 1, \ldots, K_{4} \,,
\label{BetheYang3}
\ee
and the BAEs (\ref{BAEsu2}), (\ref{BAEsu22}) become
\be
\prod_{i=1}^{K_{4}} \frac{\alpha(\xi_{k})-\alpha(p_{4,i})+\frac{i}{2}}
{\alpha(\xi_{k})-\alpha(p_{4,i})-\frac{i}{2}}
&=& \prod_{j=1 \atop j\ne k}^{m_{0}}
\frac{\alpha(\xi_{k})-\alpha(\xi_{j})+i}
{\alpha(\xi_{k})-\alpha(\xi_{j})-i}\,, \qquad k = 1\,, \ldots \,, m_{0} \,, \non \\
\prod_{i=1}^{K_{4}} \frac{1 - \frac{1}{x_{1,j} x^{-}_{4,i}}}
{1 - \frac{1}{x_{1,j} x^{+}_{4,i}}}
&=& \prod_{l=1}^{K_{2}}
\frac{u_{1,j} - u_{2,l} + \frac{i}{2}}
{u_{1,j} - u_{2,l} - \frac{i}{2}} \,, \qquad j = 1, \ldots, K_{1} \,,\non \\
\prod_{i=1}^{K_{4}} \frac{x_{3,j}-x^{-}_{4,i}}
{x_{3,j}-x^{+}_{4,i}}
&=& \prod_{l=1}^{K_{2}}
\frac{u_{3,j} - u_{2,l} + \frac{i}{2}}
{u_{3,j} - u_{2,l} - \frac{i}{2}} \,, \qquad j = 1, \ldots, K_{3} \,,\non \\
\prod_{j=1}^{K_{1}}\frac{u_{2,l} - u_{1,j}+\frac{i}{2}}
{u_{2,l} - u_{1,j}-\frac{i}{2}}
\prod_{j=1}^{K_{3}}\frac{u_{2,l} - u_{3,j}+\frac{i}{2}}
{u_{2,l} - u_{3,j}-\frac{i}{2}}
&=&
\prod_{k=1 \atop k\ne l}^{K_{2}}
\frac{u_{2,l} - u_{2,k} +i}
{u_{2,l} - u_{2,k} -i} \,, \qquad l = 1, \ldots, K_{2} \,,
\label{BAEgen}
\ee
respectively. Eqs. (\ref{BetheYang3}),  (\ref{BAEgen}) constitute our 
result for the all-loop BAEs corresponding to the $S$-matrix 
(\ref{Smatrix}),  (\ref{su2S}).

The weak-coupling limit corresponds to \cite{GV}
\be
x \rightarrow \frac{u}{g}\,, \qquad x^{\pm} \rightarrow \frac{1}{g}\left(u
\pm \frac{i}{2}\right) \,,
\ee
with $g\rightarrow 0$ and $u$ finite. Recalling (\ref{weak}), we obtain
\be
\left(\frac{u_{4,k}+\frac{i}{2}}{u_{4,k}-\frac{i}{2}} \right)^{L} &=&
\prod_{i=1}^{K_{4}}\left\{ S_{0}(p_{4,k}, p_{4,i})\,
\left(p_{4,k} -p_{4,i} + i\right)
\left(\frac{u_{4,k} - u_{4,i}+i}{u_{4,k} - u_{4,i}-i} \right)
\right\} \non \\
& & \times \prod_{j=1}^{m_{0}}\frac{\xi_{j} -p_{4,k} + \frac{i}{2}}
{\xi_{j} -p_{4,k} - \frac{i}{2}}
\prod_{j=1}^{K_{3}}
\frac{u_{4,k}-u_{3,j}-\frac{i}{2}}
{u_{4,k}-u_{3,j}+\frac{i}{2}}
\,, \quad k = 1, \ldots, K_{4} \,, \non \\
1 &=& \prod_{j=1 \atop j\ne k}^{m_{0}}
\frac{\xi_{k}-\xi_{j}+i}
{\xi_{k}-\xi_{j}-i}
\prod_{i=1}^{K_{4}} \frac{\xi_{k}-p_{4,i}-\frac{i}{2}}
{\xi_{k}-p_{4,i}+\frac{i}{2}}\,, \qquad k = 1\,, \ldots \,, m_{0} \,, \non 
\ee
\be 
1 &=& \prod_{l=1}^{K_{2}}
\frac{u_{1,j} - u_{2,l} + \frac{i}{2}}
{u_{1,j} - u_{2,l} - \frac{i}{2}} \,, \qquad j = 1, \ldots, K_{1} \,,\non \\
1 &=& \prod_{l=1}^{K_{2}}
\frac{u_{3,j} - u_{2,l} + \frac{i}{2}}
{u_{3,j} - u_{2,l} - \frac{i}{2}}
\prod_{i=1}^{K_{4}} \frac{u_{3,j}-u_{4,i}-\frac{i}{2}}
{u_{3,j}-u_{4,i}+\frac{i}{2}} \,, \qquad j = 1, \ldots, K_{3} \,, 
\label{BAEgenweak} \\
1 &=&
\prod_{k=1 \atop k\ne l}^{K_{2}}
\frac{u_{2,l} - u_{2,k} -i}
{u_{2,l} - u_{2,k} +i}
\prod_{j=1}^{K_{1}}\frac{u_{2,l} - u_{1,j}+\frac{i}{2}}
{u_{2,l} - u_{1,j}-\frac{i}{2}}
\prod_{j=1}^{K_{3}}\frac{u_{2,l} - u_{3,j}+\frac{i}{2}}
{u_{2,l} - u_{3,j}-\frac{i}{2}}
\,, \qquad l = 1, \ldots, K_{2} \,, \non
\ee
where we have defined
\be
L = -L'+\frac{K_{4}+K_{1}-K_{3}}{2} \,,
\ee
and used
\be
e^{i p_{4,k}} = \frac{u_{4,k}+\frac{i}{2}}{u_{4,k}-\frac{i}{2}} \,.
\ee
Evidently, regardless of the choice of scalar factor
$S_{0}(p_{1},p_{2})$, the set of BAEs (\ref{BAEgenweak}) does not
completely match any of the equivalent sets of BAEs proposed by
Minahan and Zarembo \cite{MZ}.  In particular, while the latter have
two ``massive'' nodes, the former has only one.  We would have
obtained a similar result had we chosen the flavor $S$-matrix to be of
the XYZ chain/8-vertex model form rather than (\ref{su2S}). We 
conclude that an $S$-matrix of the form (\ref{Smatrix}) is not 
consistent with the perturbative BAEs \cite{MZ}.

\section{Discussion}\label{sec:discussion}

We have considered an alternative $S$-matrix for ${\cal N}=6$ CS which
is symmetric under $SU(2|2)$.  In contrast with our original proposal
\cite{AN}, this $S$-matrix has the tensor product form
(\ref{Smatrix}); and it has not only an $SU(2|2)$ part, but also a
nontrivial flavor part which allows for reflection in $A-B$
scattering.  Although we have not proved that this tensor product
structure is the only possible way of introducing reflection while
both maintaining integrability and respecting the system's symmetry,
we have not found any other.  We have argued that such an $S$-matrix
is not consistent with the perturbative BAEs \cite{MZ}.  This gives
increased confidence in our original proposal \cite{AN}.  Further
support for the proposal \cite{AN} has recently been found in
computations of finite-size corrections to the dispersion relation of
giant magnons \cite{BF, LS}, and in the direct coordinate Bethe ansatz
computation of the two-loop scalar-sector $S$-matrix \cite{AN2}.

\section*{Acknowledgments}
We thank K. Zarembo for raising the question which we address, and for
helpful discussions.  We are also grateful to the Referee for comments
on an earlier version of this paper.  This work was supported in part
by KRF-2007-313-C00150 (CA) and by the National Science Foundation
under Grants PHY-0244261 and PHY-0554821 (RN).


\begin{thebibliography}{99}

\bibitem{ABJM}
O. Aharony, O. Bergman, D.~L. Jafferis and J. Maldacena,
``${\cal N}=6$ superconformal Chern-Simons-matter theories, M2-branes and their
gravity duals,''
{\it JHEP} {\bf 0810}, 091 (2008)
[arXiv:0806.1218].

\bibitem{MZ}
J.~A. Minahan and K. Zarembo,
``The Bethe ansatz for superconformal Chern-Simons,''
{\it JHEP} {\bf 0809}, 040  (2008)
[arXiv:0806.3951].

\bibitem{BR}
D.~Bak and S.~J.~Rey,
``Integrable Spin Chain in Superconformal Chern-Simons Theory,''
{\it JHEP} {\bf 0810}, 053 (2008)
[arXiv:0807.2063].

\bibitem{GV} N. Gromov and P. Vieira,
``The all loop AdS4/CFT3 Bethe ansatz,'' 
{\it JHEP} {\bf 0901}, 016 (2009)
[arXiv:0807.0777].

\bibitem{MR}
T.~McLoughlin and R.~Roiban,
``Spinning strings at one-loop in $AdS_4 \times P^3$,''
{\it JHEP} {\bf 0812}, 101 (2008)
[arXiv:0807.3965].

\bibitem{AAB}
L.~F.~Alday, G.~Arutyunov and D.~Bykov,
``Semiclassical Quantization of Spinning Strings in $AdS_4 \times CP^3$,''
{\it JHEP} {\bf 0811}, 089 (2008) 
[arXiv:0807.4400].

\bibitem{Kr}
C.~Krishnan,
``AdS4/CFT3 at One Loop,''
{\it JHEP} {\bf 0809}, 092 (2008)
[arXiv:0807.4561].

\bibitem{MRT}
T.~McLoughlin, R.~Roiban and A.~A.~Tseytlin,
``Quantum spinning strings in $AdS_4 \times CP^3$: testing the Bethe Ansatz proposal,''
 {\it JHEP} {\bf 0811}, 069 (2008) 
[arXiv:0809.4038].

\bibitem{GM}
N.~Gromov and V.~Mikhaylov,
``Comment on the Scaling Function in $AdS4 \times CP3$,''
[arXiv:0807.4897].

\bibitem{NT}
T. Nishioka and T. Takayanagi,
``On Type IIA Penrose Limit and ${\cal N}=6$ Chern-Simons Theories,''
{\it JHEP} {\bf 0808}, 001  (2008)
[arXiv:0806.3391].

\bibitem{GGY}
D. Gaiotto, S. Giombi and X. Yin,
``Spin Chains in ${\cal N}=6$ Superconformal Chern-Simons-Matter Theory,''
[arXiv:0806.4589].

\bibitem{GHO}
G.~Grignani, T.~Harmark and M.~Orselli,
``The SU(2) x SU(2) sector in the string dual of ${\cal N}=6$ superconformal
Chern-Simons theory,''
[arXiv:0806.4959].

\bibitem{AN} 
C.~Ahn and R.~I.~Nepomechie,
``${\cal N}=6$ super Chern-Simons theory $S$-matrix and all-loop Bethe ansatz
equations,''
{\it JHEP} {\bf 0809}, 010 (2008)
[arXiv:0807.1924].

\bibitem{BT}
D.~Berenstein and D.~Trancanelli,
``Three-dimensional ${\cal N}=6$  SCFT's and their membrane dynamics,''
[arXiv:0808.2503].

\bibitem{St}
M.~Staudacher,
``The factorized S-matrix of CFT/AdS,''
{\it JHEP} {\bf 0505}, 054 (2005)
[arXiv:hep-th/0412188].

\bibitem{Be1}
N. Beisert,
``The $su(2|2)$ dynamic $S$-matrix,''
{\it Adv. Theor. Math. Phys.}  {\bf 12}, 945 (2008)
[arXiv:hep-th/0511082];\\
N. Beisert,
``The Analytic Bethe Ansatz for a Chain with Centrally Extended
$su(2|2)$ Symmetry,''
{\it J. Stat. Mech.}  {\bf 0701}, P017 (2007)
[arXiv:nlin/0610017].

\bibitem{MM}
M.J. Martins and C.S. Melo,
``The Bethe ansatz approach for factorizable centrally extended
$S$-matrices,''
{\it Nucl. Phys.} {\bf B 785}, 246 (2007)
[arXiv:hep-th/0703086].

\bibitem{AFZ}
G. Arutyunov, S. Frolov and M. Zamaklar,
`The Zamolodchikov-Faddeev algebra for $AdS_{5} \times S^{5}$ superstring,''
{\it JHEP} {\bf 0704}, 002 (2007)
[arXiv:hep-th/0612229].

\bibitem{Be2}
N. Beisert, ``Integrability in AdS/CFT,''
lecture at the workshop {\it Strong Fields, Integrability
and Strings}, Newton Institute (2007).

\bibitem{product}
C. Ahn, D. Bernard and A. LeClair, 
``Fractional supersymmetries in perturbed coset CFTs and integrable 
soliton theory,''
{\it Nucl. Phys.} {\bf B346}, 409 (1990);\\
N. Reshetikhin,
``$S$-matrices in integrable models of isotropic magnetic chains. I,''
{\it J. Phys.} {\bf A24}, 3299 (1991).

\bibitem{BF}
D.~Bombardelli and D.~Fioravanti,
``Finite-Size Corrections of the $\mathbb{CP}^3$ Giant Magnons: the
L\"{u}scher terms,''
[arXiv:0810.0704].

\bibitem{LS}
T.~Lukowski and O.~O.~Sax,
``Finite size giant magnons in the $SU(2) \times SU(2)$ sector of 
$AdS_4 \times CP^3$,''
{\it JHEP} {\bf 0812}, 073 (2008)
[arXiv:0810.1246].

\bibitem{AN2} 
C.~Ahn and R.~I.~Nepomechie,
``Two-loop test of the ${\cal N}=6$ Chern-Simons theory $S$-matrix,''
[arXiv:0901.3334].

\end{thebibliography}
\end{document}